# A momentum-dependent perspective on quasiparticle interference in $Bi_2Sr_2CaCu_2O_{8+\delta}$

I. M. Vishik,[1,2] B. Moritz,[2] E. A. Nowadnick,[1,2] W. S. Lee,[1,2] K. Tanaka,[3] T. Sasagawa,[4] T. Fujii,[5] T. P. Devereaux,[2] and Z. X. Shen[1,2]

1. Geballe Laboratory for Advanced Materials,
Departments of Physics and Applied Physics, Stanford University, CA 94305
2. Stanford Institute for Materials and Energy Sciences,
SLAC National Accelerator Laboratory,
2575 Sand Hill Road, Menlo Park, CA 94025
3. Department of Physics, Osaka University, Osaka, Japan
4. Materials and Structures Laboratory,
Tokyo Institute of Technology, Tokyo, Japan
5. Cryogenic Center, University of Tokyo, Tokyo, Japan

**Angle Resolved Photoemission Spectroscopy (ARPES) probes the momentum-space electronic structure of materials, and provides invaluable information about the high-temperature superconducting cuprates [1]. Likewise, cuprates' real-space, inhomogeneous electronic structure is elucidated by Scanning Tunneling Spectroscopy (STS). Recently, STS has exploited quasiparticle interference (QPI) – wave-like electrons scattering off impurities to produce periodic interference patterns – to infer properties of the QP in momentum-space. Surprisingly, some interference peaks in $Bi_2Sr_2CaCu_2O_{8+\delta}$ (Bi-2212) are absent beyond the antiferromagnetic (AF) zone boundary, implying the dominance of particular scattering process [2]. Here, we show that ARPES sees no evidence of quasiparticle (QP) extinction: QP-like peaks are measured everywhere on the Fermi**

**surface, evolving smoothly across the AF zone boundary. This apparent contradiction stems from different natures of single-particle (ARPES) and two-particle (STS) processes underlying these probes. Using a simple model, we demonstrate extinction of QPI without implying the loss of QP beyond the AF zone boundary.**

Recently, STS has been used to infer momentum-space information via the Fourier-transform (FT) of the local density of states (LDOS), $\rho(\mathbf{r},\omega)$.[3-6] Conventionally, a superconductor has well-defined momentum-eigenstates (i.e. Bogoliubov QP), so $\rho(\mathbf{r},\omega)$ is spatially homogenous. However, the cuprates are intrinsically inhomogeneous,[7] and scattering off these impurities mixes momentum eigenstates. QPI manifests itself as a spatial modulation of $\rho(\mathbf{r},\omega)$ with well-defined $\mathbf{q}$, appearing in the FT, $\rho(\mathbf{q},\omega)$. QPI experiments are interpreted in terms of the octet model, [3-6] positing that wavevectors $\mathbf{q_{1-7}}$ connecting the ends of "banana-shaped" contours of constant energy (CCEs) dominate $\rho(\mathbf{q},\omega)$. Dispersing $\mathbf{q}$ are associated with coherent superconducting quasiparticles, and the evolution of $\mathbf{q_{1-7}}$ as a function of bias-voltage is used to infer the Fermi surface (FS) and the magnitude of the *d*-wave superconducting gap. QPI experiments have found that the intensity of some of the peaks in $\rho(\mathbf{q},\omega)$ vanishes approaching the diagonal line between $(0,\pi)$ and $(\pi,0)$ (corresponding to the AF zone boundary).[5, 6] These results lead naturally to speculation about the disappearance of QPI after AF zone and possible extinction of QP themselves near the boundary of Brillouin zone (antinodal region) due to strong scattering near the extinction point.

Notably, ARPES has long observed antinodal QP below $T_c$ in Bi-2212 over a wide doping range (p > 0.08), although this point has not been a focus in the literature.[8-10]. In Fig. 1, we report data on single crystals of Bi-2212 at four dopings spanning much of the same doping range as recent STS studies,[5] going from slightly underdoped ($T_c$=92K, denoted UD92) to substantially underdoped (UD50). There are sharp peaks in all the energy distribution curves (EDCs) at low binding energy, strongly arguing against extinction in the single-particle spectrum. We note that for underdoped Bi-2212, the antinodal QP has been shown to scale with the superfluid density [2,8] and $T_c$. In addition, since the pseudogap remains near the antinode even for $T>T_c$, in this region of momentum-space, the emergence of a QP peak below $T_c$ [11] (Fig. 1(f), inset)_is the primary hallmark of superconductivity. In Fig. 1(f), we also show the locus of points on the FS as determined by FT-STS for UD74,[5] and a comparison to the ARPES FS on a sample with similar doping ($T_c$ = 75 K). Both techniques yield a similar FS, but FT-STS only observes a dispersing interference pattern over a limited momentum range terminating at the AF zone boundary, whereas ARPES sees QP all over the entire FS.

To assess whether there is any anomalous scattering near the AF zone boundary, we have fit the peaks in Fig. 1. Although many factors contribute to the amplitude and width of the QP peaks, including matrix element and $k_z$ effects,[12] qualitative conclusions about the single-particle scattering rate can be deduced by analyzing the momentum dependence of the peak width. Symmetrized EDCs were fit to a spectral function with an energy-dependent scattering rate, $\Gamma(\omega)=\alpha\omega$, convolved with the experimental energy resolution. A similar model was found to provide robust fit for the spatially-inhomogeneous STS conductances seen in Bi-2212.[13] At $\mathbf{k}=\mathbf{k}_F$, the only free parameters are the scattering prefactor, $\alpha$ and the gap energy, $\Delta(\mathbf{k})$; together, these define a characteristic scattering rate for each point along the FS, $\Gamma_2^*(\mathbf{k})=$     $(\mathbf{k})$. Figure 2 shows $\Gamma_2^*$ as a function of FS angle for UD75 and UD92, plotted together with the fitted gap.

Near the antinode, the peak width is smaller than the gap, implying that the peaks are QP-like in this region of interest. For UD92, the QP width changes little going from the node to the antinode, a result which is similar to earlier work on overdoped samples.[14] The UD75 scattering rate shows stronger momentum dependence, but the overall variation is still only a factor of three. For both dopings, there is no anomaly in the scattering at the AF zone boundary. This rules out proposals for QP extinction which invoke a sudden increase in single-particle scattering at the AF zone boundary.[5]

What could plausibly explain the apparent contradiction between ARPES and FT-STS results? One intriguing fact is that not all QPI wavevectors in $\rho(\mathbf{q},\omega)$ vanish across the AF zone boundary; $\mathbf{q}_{1,4,5}$ survive while the others fall below the FT-STS noise floor. Here we study the effects of QP scattering from impurities based on a weak-coupling approach. While this neglects the large, relevant, spatial inhomogeneity in the LDOS observed in STS,[7] it places a simple focus on the differences between measurements, contrasting electron removal spectra in ARPES against two-particle QPI mechanisms in FT-STS.

Based on ARPES results, the electron propagators are described by the BCS Green's function $G$ in the superconducting state, and the non-uniform part of $\rho(\mathbf{q},\omega)$, is determined by the momentum-dependent $\hat{T}$-matrix:[15]

$$\delta\rho(\mathbf{q},\omega) = \frac{-1}{\pi} \operatorname{Im} \sum_{\mathbf{p}} [\hat{G}(\mathbf{p},\omega)\hat{T}(\mathbf{p},\mathbf{p}+\mathbf{q})\hat{G}(\mathbf{p}+\mathbf{q},\omega)]_{11}$$

Where $\hat{G}(\mathbf{p},\omega) = 1/[\omega\hat{\tau}_0 - \xi(\mathbf{p})\hat{\tau}_3 - \Delta(\mathbf{p})\hat{\tau}_1]$ in terms of Pauli matrices $\hat{\tau}_{0,1,3}$. $\xi(\mathbf{p})$ is the band structure, and $\Delta(\mathbf{p}) = \Delta_0[\cos(k_x a) - \cos(k_y a)]/2$ is the $d$-wave superconducting gap.

Contributions to the $\hat{T}$-matrix can be classified according to modification of electron parameters: conventional impurity scattering enters in the $\hat{\tau}_3$ channel, while local superconducting gap modification occurs in the $\hat{\tau}_1$ channel.

Here, a single impurity at site (0,0) locally modifies hopping and the *d*-wave superconducting gap through an impurity contribution to the Hamiltonian: $H = \sum_r \Psi_r^\dagger [\hat{\tau}_3 \delta t(r) + \hat{\tau}_1 \delta\Delta(r)] \Psi_0 + h.c.$, where spin indices have been suppressed. The hopping and d-wave gap modulations δt(r) and δΔ(r) are proportional to $[\delta(r-ax) \pm \delta(r-ay) + \delta(r+ax) \pm \delta(r+ay)]$ where *a* is the square lattice spacing with the upper (lower) sign for hopping (d-wave gap) modulation. Given this form of the impurity Hamiltonian, the Fourier transform, $\hat{T}(\mathbf{p},\mathbf{p}+\mathbf{q})$, has a simple momentum space form: $\hat{\tau}_{3,1}[\cos(p_x) \pm \cos(p_y) + \cos(p_x + q_x) \pm \cos(p_y + q_y)]$ for hopping and *d*-wave gap modulated scattering, respectively. In the $\hat{\tau}_1$ channel (gap modulation) scattering vanishes between points with opposite order parameter phase. Thus $\mathbf{q}_{2,3,6,7}$ vanish while $\mathbf{q}_{1,4,5}$ persist. In the $\hat{\tau}_3$ channel (hopping modulation) scattering between equivalent points leads to no such cancellation, but the momentum dependence of the $\hat{T}$-matrix implies a loss of intensity for **q**-vectors connecting points **p** and **p+q** that both lie on the AF zone boundary. Thus the disappearance of QPI peaks associated with $\mathbf{q}_{2,3,6,7}$ at the AF zone boundary for the types of disorder considered here may be associated simply with the momentum dependence of the $\hat{T}$-matrix, rather than implying the 'extinction' of QP. This also may reconcile the observation that the QPI peaks vanish even at large dopings (p≈0.19), a regime with a diminished influence from the AF Mott insulating state.

While the analysis presented above for a single impurity scatterer embedded in an infinite host provides a simple view of the momentum dependence of the $\hat{T}$-matrix and the plausible loss of QPI peak intensity, it cannot describe the behavior of FT-STS QPI peak intensity at a general bias voltage or the implications associated with the FT of the *Z*-map rather than the LDOS itself. A more qualitative, and quantitative, comparison with FT-STS results from considering the effects of an extended 'patch' impurity embedded in a finite-size, periodic host with the $\hat{T}$-matrix determined self-consistently. Finite size effects are partially mitigated by smoothing the impurity with a Gaussian envelope intended to decrease the influence of the impurity on electronic parameters with the distance from the center of the patch. In addition, the QPI intensity is determined from a FT of the *Z*-map, a ratio of the LDOS at positive and negative bias, as done in FT-STS experiments.[4-5]. More details of the calculation including the exact patch geometry and modulation parameters can be found in the supplementary material accompanying this letter.

The panels in Fig. 3 show results for 5% (3c-d), 10% (3e-f), and 15% (3g-h) hole doping. The dashed line in each panel indicates the energy associated with the AF zone boundary. Note the general trends agree with the initial analysis from a single impurity scatterer. Intensities for peaks $q_{2,3,6,7}$, shown in the left panels, initially rise moving away from the nodal point and then begin to fall approaching the AF zone boundary. Intensity for QPI peaks $q_{1,5}$ is small initially and rises abruptly as the bias voltage approaches and crosses the energy associated with the AF zone boundary, both trends in agreement with the single impurity analysis. This behavior is in qualitative agreement with the experimental intensity profiles as a function of energy for all these QPI peaks.[5] While there are general quantitative changes with band structure parameters, impurity 'patch' size and shape, and degree of modulation, the qualitative agreement with the single impurity analysis and experimental findings remains robust.

We have presented systematic ARPES data that demonstrate the ubiquity of QPs around the FS for a wide doping range including the heavily underdoped regime. Thus, the disappearance of QPI across the AF zone boundary is not due to the extinction of QP near the antinodal region. Instead, as suggested by our impurity model calculations, momentum-dependent impurity scattering reconciles the contradiction between one-particle (ARPES) and two-particle (FT-STS) observations.

We thank Profs. N. Nagaosa, and J. Zaanen for helpful discussions. SSRL is operated by the DOE Office of Basic Energy Science, Division of Chemical Science and Material Science. This work is supported by DOE Office of Science, Division of Materials Science, with contracts DE-FG03-01ER45929-A001, DE-AC02-76SF00515, and NSF grant DMR-0604701.

**Methods**

ARPES measurements were performed at Beamline 5-4 of Stanford Synchrotron Radiation Lightsource (SSRL) using a Scienta R4000 electron analyzer. Samples were cleaved *in-situ* at a pressure better than $5\times10^{-11}$ torr. Measurements were performed at 10K with an energy resolution of 8meV. UD50 was measured with 19eV photons in the second Brillouin zone (BZ) with cuts parallel to ΓY. The other samples were measured with 22.7eV photons in the first BZ with cuts parallel to ΓM.

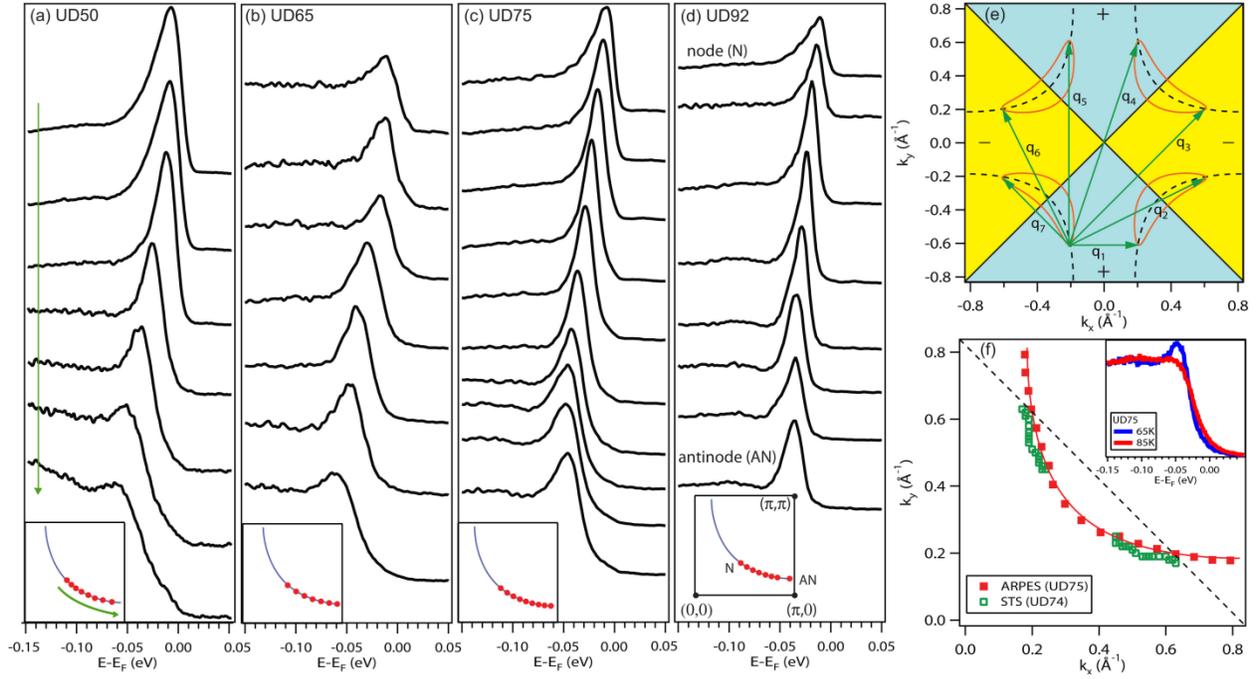

**FIG. 1: QP in ARPES data.** (a)-(d) EDCs at $\mathbf{k}_F$ from the node (top) to the antinode (bottom) for four different dopings. Insets show where the cuts intersect the FS. The reduction of the peak intensity and increase in peak width near the node seen in (b)-(d) stems from matrix element and momentum resolution effects which are described in the Supplementary Materials. (e) QPI wavevectors within the octet model. When a $d$-wave gap opens on the FS (dotted line), QPI measurements are dominated by scattering between the ends of CCEs (red solid lines).[5] Blue and yellow regions represents $\Delta(\mathbf{k})>0$ and $\Delta(\mathbf{k})<0$, respectively. (f) By measuring how $\mathbf{q_1}$-$\mathbf{q_7}$ disperse, the ends of the CCEs can be tracked as a function of energy to map out the FS. QPI experiments suggested that the locus of coherent QPs (open green squares) terminates at the line connecting $(0,\pi)$ and $(\pi,0)$ (dotted line).[5] Meanwhile, ARPES shows that the locus of coherent QPs (red filled squares) extends all the way to the antinode for a similar doping. Inset: UD75 EDCs at the antinode measured at 85K (red) and 65K (blue), indicating that the peak sets in near $T_c$.

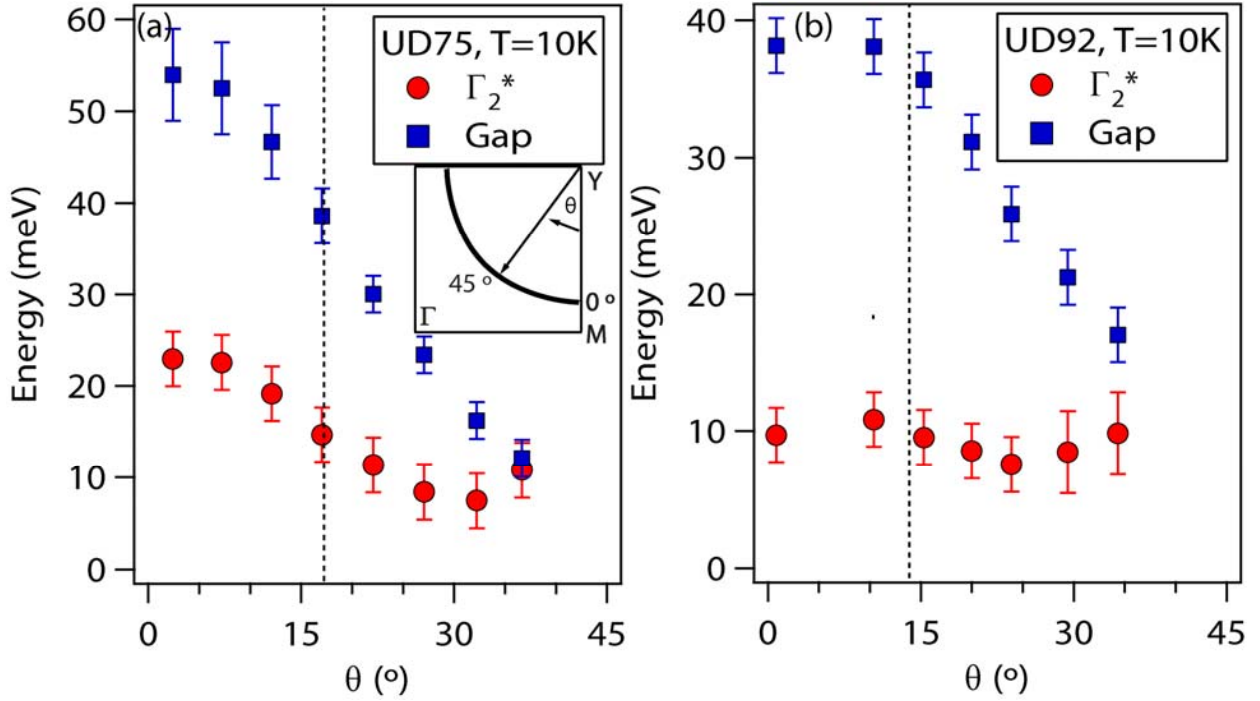

**FIG. 2: Scattering-rate fits.** Symmetrized EDCs at $k_F$ for UD75 (a) and UD92 (b) were fit to a model with an energy-dependent scattering rate, $\Gamma = \alpha\omega$, convolved with the experimental energy resolution. The characteristic scattering rate, $\Gamma_2^* = \alpha\Delta(\mathbf{k})$ is plotted as a function of FS angle along with the fitted $\Delta(\mathbf{k})$. Error bars for $\Gamma_2^*$ are $3\sigma$ confidence error bars from the fitting, and error bars for the gap are described in Ref. [10]. There is no evidence that the scattering rate diverges near the AF zone boundary for either doping. Additionally, the peak width in the antinodal region is smaller than the gap, indicating that the peaks are QP-like.

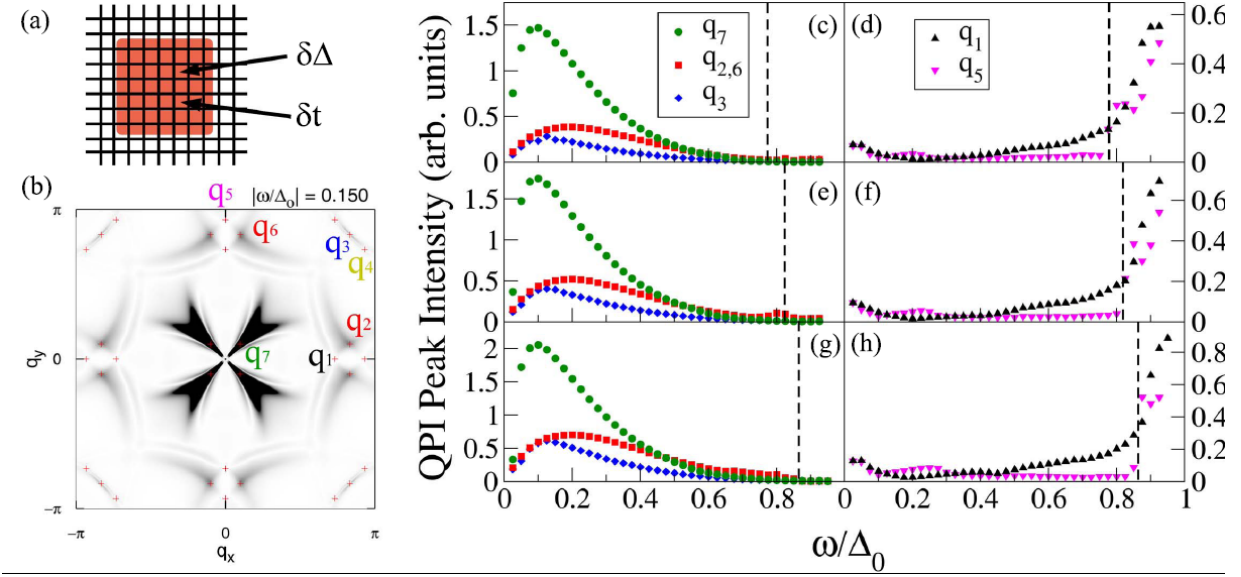

**FIG. 3: Patch Calculation.** (a) Sketch of the patch used for the calculations. (b) *Z*-map for $|\omega/\Delta_0|=0.15$. The qp wavevectors $q_1$-$q_7$ are indicated. QPI peak intensities determined from the Z-map for (c)-(d) 5%, (e)-(f) 10%, and (g)-(h) 15% hole doping. Panels (c), (e), and (g) show the intensity for wavevectors $q_7$, $q_{2,6}$, and $q_3$. The dashed line in each panel indicates the energy at which the tips of the CCE cross the AF zone boundary, the point at which the peak intensities associated with $q_{2,3,6,7}$ vanish. Panels (d), (f), and (h) show the intensity for wavevectors $q_1$ and $q_5$ which rise in intensity upon approaching the AF zone boundary.